\documentclass[manuscript]{aastex}
\usepackage{emulateapj5}
\usepackage{apjfonts}
 1

\slugcomment{The Astrophysical Journal, draft version \today}

\def\gtsima{$\; \buildrel > \over \sim \;$}
\def\ltsima{$\; \buildrel < \over \sim \;$}
\def\gsim{\lower.7ex\hbox{\gtsima}}
\def\lsim{\lower.7ex\hbox{\ltsima}}
\def\simgt{\lower.7ex\hbox{\gtsima}}
\def\simlt{\lower.7ex\hbox{\ltsima}}
\def\la{\lsim}

\def\HI{\ifmmode \hbox{\scriptsize H\kern0.5pt{\footnotesize\sc i}}\else H\kern1pt{\small I}\fi}
\def\Halpha{H$\alpha$}
\def\mlstar{\ifmmode\Upsilon_{\!\!*}\else$\Upsilon_{\!\!*}$\fi}
\def\mlstarR{\ifmmode\Upsilon_{\!\!*}^R\else$\Upsilon_{\!\!*}^R$\fi}
\def\kms{\ifmmode\mathrm{~km~}\mathrm{s}^{-1}\else km s$^{-1}$\fi}
\def\cm2{cm$^{-2}$}
\def\pc2{pc$^{-2}$}
\def\pc3{pc$^{-3}$}

\def\lab{\rlap{\raise2pt\hbox{$<$}}\lower2.5pt\hbox{$\sim$}}
\def\gab{\rlap{\raise2pt\hbox{$>$}}\lower2.5pt\hbox{$\sim$}}
\def\ds0{\ifmmode d_\mathrm{S0}\else $d_\mathrm{S0}$\fi}

\newcommand{\tskip}{\omit\tablevspace{1pt}}
\shorttitle{Stellar motions in NGC 4650A}
\shortauthors{Swaters \& Rubin}

\begin{document}

\title{Stellar Motions in the Polar Ring Galaxy NGC 4650A}

\author{Rob A. Swaters}
\affil{Department of Physics and Astronomy, Johns Hopkins University,
  3400 N. Charles Str., Baltimore, MD 21218, and Space Telescope
  Science Institute, 3700 San Martin Dr., Baltimore, MD 21218;
  swaters@pha.jhu.edu}
\medskip

\and

\author{Vera C. Rubin}
\affil{Department of Terrestrial Magnetism, Carnegie Institution of
Washington, 5241 Broad Branch Rd NW, Washington DC 20015;
rubin@dtm.ciw.edu}

\begin{abstract}

  We present the first measurement of the stellar kinematics in the
  polar ring of NGC~4650A. There is well defined rotation, with the
  stars and gas rotating in the same direction, and with similar
  amplitude.  The gaseous and stellar kinematics suggest an
  approximately flat rotation curve, providing further support for the
  hypothesis that the polar material resides in a disk rather than in
  a ring. The kinematics of the emission line gas at and near the
  center of the S0 suggests that the polar disk lacks a central hole.
  We have not detected evidence for two, equal mass, counterrotating
  stellar polar streams, as is predicted in the resonance levitation
  model proposed by Tremaine \& Yu.  A merger seems the most likely
  explanation for the structure and kinematics of NGC~4650A.

\end{abstract}


\section{Introduction}

Although polar ring galaxies constitute only a small fraction of the
galaxy population, these objects are interesting because they offer a
unique insight into the formation and evolution of galaxies. For
example, understanding the kinematics of polar rings could ultimately
reveal the three-dimensional shape of their dark matter halos
(Whitmore, McElroy, \& Schweizer 1987; Sackett et al.\ 1994; Combes \&
Arnaboldi 1996). They may also provide information on mergers of
galaxies. The conventional model for polar ring formation (Schweizer,
Whitmore, \& Rubin\ 1983; Schechter, Ulrick, \& Boksenberg\ 1984)
attributes ring formation to gas capture after the central galaxy has
formed. Bekki\ (1998) has suggested a different model in which a polar
ring galaxy is formed in a dissipative polar merger of two comparable
mass disk galaxies with small relative velocity.

\begin{figure*}
\begin{center}
\resizebox{0.8\hsize}{!}{\includegraphics{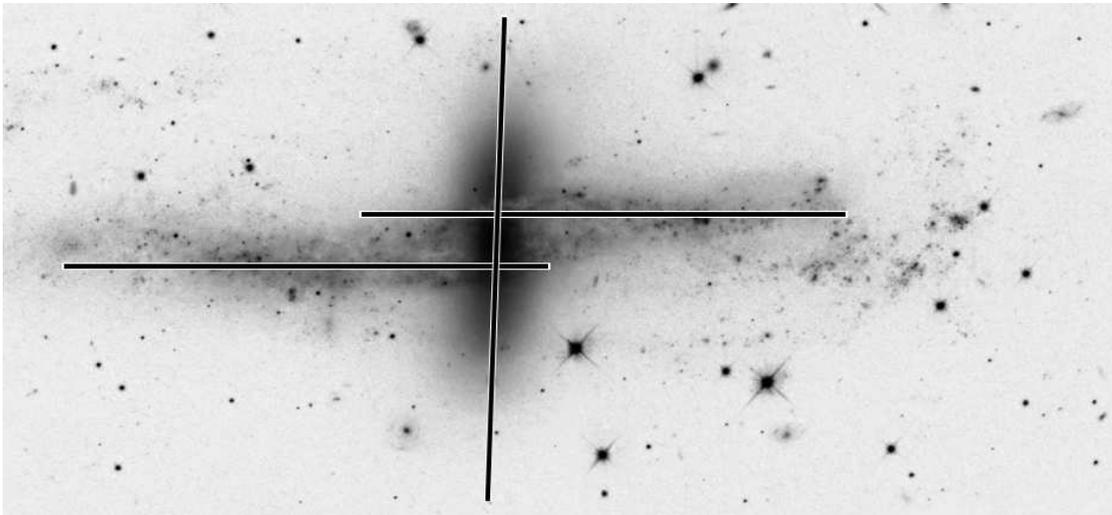}}
\caption{Image of NGC~4650A from the Hubble Space Telescope Heritage
program, with the slit positions overlayed. The image size is $2.7'$
by $1.3'$. North is $110^\circ$ and east is $200^\circ$
counterclockwise from the top. \label{figslitpos}}
\end{center}
\end{figure*}

Recently, Tremaine \& Yu (2000) have offered an alternative model for
polar ring formation that requires no accretion or merger. If a galaxy
lies in a symmetry plane of a triaxial dark halo with an initially
retrograde pattern speed that slowly tends to zero (e.g., as the halo
acquires dark matter by infall), disk stars are trapped at a
resonance when the rate of precession of a slightly tilted orbit
matches the pattern speed of the triaxial halo.  As the resonance
slowly sweeps outward past the stars, they are levitated into polar
orbits.  The model predicts that a polar ring formed in this way will
contain two equal mass, counterrotating disk streams. Note that, if
the halo pattern speed were to increase through zero to prograde, the
stellar orbits could become retrograde in the disk, and form a
counterrotating disk stream.  Thus, the Tremaine \& Yu (2000) model
provides an interesting mechanism that can explain both polar ring and
counterrotating {\it disk} galaxies without a merger (see also Evans
\& Collett 1994). Because in the resonance levitation model the polar
ring contains counterrotating stellar populations, whereas in merger
models the ring stars would most likely be rotating in one direction,
the stellar kinematics of a polar ring provide a means to discriminate
between these formation scenarios.

A prime candidate for the measurement of the stellar velocities in a
polar ring is NGC~4650A. It consists of a central S0 galaxy,
surrounded by a prominent polar ring. With an adopted Local Group
centric velocity of V$_{LG}$=2635 km s$^{-1}$, and a Hubble constant
of 70 km s$^{-1}$ Mpc$^{-1}$, NGC~4650A is at a distance of 38~Mpc. At
this distance, $1''$=184 pc. Because of its relative proximity,
NGC~4650A has been extensively studied (see e.g., Gallagher et al.\
2002 and references therein).  Observations at 21-cm (Arnaboldi et
al.\ 1997) revealed a spatially extended \HI\ disk coincident with the
ring, leading these authors to suggest that NGC 4650A is more
appropriately called a polar disk galaxy (see also Wakamatsu 1993).
Recent observations taken with the Hubble Space Telescope as part of
the Heritage project reveal a complex, warped structure of the polar
disk (Gallagher et al.\ 2002; Iodice et al.\ 2002).

In this Letter, we report stellar and gas velocities in both the S0
and the ring of NGC 4650A.  We believe that these are the first
measurement of {\it stellar} velocities in a polar ring. Our
observations support the identification of the ring as a disk, but we
continue to call it a ring, to avoid confusion with the S0 disk.

\section{Observations and Data Reduction}

We have obtained spectra of the S0 and of the ring of NGC 4650A with
the Boller and Chivens spectrograph and the Marconi CCD at the Las
Campanas Baade (6.5-m) telescope. For all observations, the slit width
was $1''$ and the slit length $72''$.  To increase the signal-to-noise
ratio, the data were binned on-chip by 4 $0.25''$ pixels in the
spatial direction.  The spectral coverage was 1600\AA, with a FWHM
resolution of 135 km s$^{-1}$ (0.78 \AA\ pix$^{-1}$).  Observational
details are provided in Table~\ref{tabobspar}.  After each science
exposure an argon wavelength calibration and an offset sky exposure
were taken.  At the beginning and end of each night template stars
with spectral types in the range G8III to K4III were observed.

The choice of slit position merits special mention.  Because of the
warp of the ring, a single slit position cannot cover both\break

\null\vspace{-1cm}\begin{center}
{\sc Table \ref{tabobspar}\\
\smallskip\hbox to\hsize{\hfil{Journal of Observations}\hfil}}
\small
\setlength{\tabcolsep}{6pt}
\begin{tabular}{llllr}
\tskip \tableline
\tableline \tskip
Date & Spectral & Integration & Position & Region \\
     & region   & time        & angle    & \\
\tableline \tskip
2001 May 29 & \Halpha &  1800s       &  155 & N ring \\
2001 May 30 & \Halpha &  1800s       &  150 & S ring \\
2002 Apr 6 &  Mg  &  12 x 1800s  &  335 & N ring \\
2002 Apr 7 &  Mg  &   7 x 1800s  &   63 & S0 major \\
2002 Apr 7 &  Mg  &   4 x 1800s  &  335 & S ring \\
\tableline
\end{tabular}
\end{center}

\noindent the center of the S0 galaxy and the brightest regions of the
N and S part of the ring.  To maximize the signal from the polar ring,
we chose to align the slits along the brightest parts of the ring. As
a result, the ring slit positions are each displaced about $4''$ from
the nucleus of the S0 galaxy. For the major axis spectrum of the S0
galaxy, a position angle of 63$^\circ$ was adopted (Whitmore et al.\
1990). The positions and orientations of the slit for each of these
pointings are indicated in Fig.~\ref{figslitpos}.

During an earlier observing run with the Baade 6.5m telescope, we
obtained H$\alpha$ emission line spectra of the N and S parts of the
ring (Table~\ref{tabobspar}).  The slit positions for these spectra
were close to those for the 2002 observations indicated in
Fig.~\ref{figslitpos}. 

The frames were bias subtracted, flat fielded, cleaned of cosmic ray
events, wavelength calibrated and combined in IRAF.  Sky subtraction
was done by combining all offset sky images for the appropriate slit
position and subtracting it, after scaling, from the corresponding
science frame.

\subsection{Derivation of the velocities}

To determine the stellar kinematics, the underlying stellar continuum
emission was first subtracted from both the galaxy and the template
star frames. Next, these continuum subtracted frames were cross
correlated by Swaters to obtain the stellar line-of-sight (LOS)
velocities.  The absorption velocities were determined by fitting
Gaussian profiles to the cross-correlation peaks (CCP), every $1''$
where the signal was strong (along almost the entire length of the S0
major axis slit position, and within $10''$ of the S0 major axis for
the ring slit positions). At other locations, the data were binned in
$4''$ intervals.  The calculated errors of the derived velocities are
the quadratic sum of the formal errors on the cross correlation and
the dispersion resulting from using different template stars. Example
CCPs are shown in the top four panels of Fig.~\ref{figccs}. The
typical width of the cross-correlation peaks is about 230 \kms. The
reduction process was not optimized to measure the stellar
dispersions.

The wavelength range of these observations also contains the [OIII]
5007, 4959, and H$\beta$ emission lines. Emission line velocities were
measured by Rubin using the customized DTM measuring program in the
software package VISTA (Rubin, Hunter, and Ford 1990). The argon
frames for each slit position were used for wavelength
calibration. Only the [OIII] 5007 line was measured because the
spectrum degraded rapidly toward the blue and the H$\beta$ emission
was located in a fairly deep absorption trough.  The 2001 \Halpha\
spectra were measured by Rubin using the same procedures as for the
[OIII] line, except that the night sky emission lines on each frame
were used for the wavelength calibration.

\section{Results}

In Fig.~\ref{figstellarkin} we plot the LOS velocities of
the polar ring stars as a function of their distances from the major
axis of the S0, \ds0. Note that $\ds0=0$ corresponds to the location
where the slit crosses the major axis, rather than the distance from
the nucleus. With the exception of the region within $|\ds0|\sim 10''$
of the S0 major axis, the stellar velocities along the ring resemble
those expected for a normal, late-type disk galaxy.

A notable feature in Fig.~\ref{figstellarkin} is the apparently
double valued velocities in the central regions.  We attribute these
to the combination of the strong contribution of light from the S0
that overwhelms the contribution from the ring stars, and the slit
positions that are offset by $4''$ from the S0 center. The radial
velocities at these positions along the S0 major axis (see
Fig.~\ref{figS0}) are indicated in Fig.~\ref{figstellarkin} by the
arrows, and match well the velocities observed in the ring spectra.

The observations reveal no evidence for equally populated
counterrotating stellar streams in the polar ring. With a velocity
difference close to 200 km s$^{-1}$ between the approaching and\break

\hskip-0.3cm\resizebox{0.99\hsize}{!}{\includegraphics{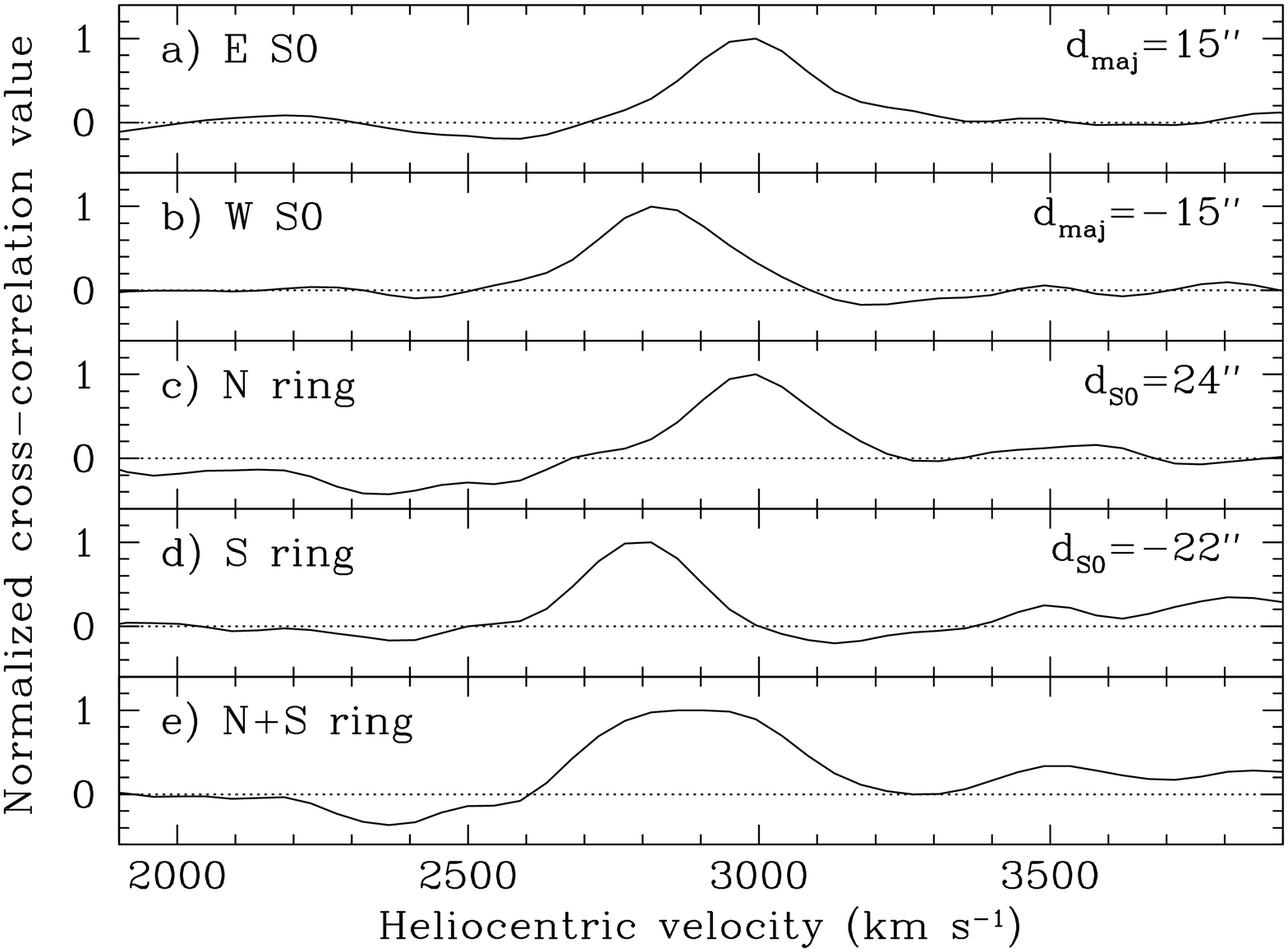}}
{\small {\sc Fig.~\ref{figccs}.}---
Example cross-correlation peaks used to determine the stellar
radial velocities. The positions of the spectra used for these
cross-correlations are given in the upper right corner of each panel
(see also Figs.~\ref{figstellarkin} and \ref{figS0}).
}
\bigskip
\addtocounter{figure}{1}

\noindent receding sides of the ring, the spectral resolution is
sufficient to detect two equal mass components. This can be seen in
Fig.~\ref{figccs}e, which shows the CCP for a model equal mass,
counterrotating stream, constructed by coadding the N and S
spectra. This model CCP, with its width of 350 \kms, is significantly
wider than the observed CCPs. Note also that the model CCP peaks near
the systemic velocity, as expected for two equal, counterrotation
streams. Thus, merely the fact that we observe significant rotation in
the ring rules against the existence of two equal mass counterrotating
streams. However, we cannot rule out the existence of a weaker,
counterrotating component, whose absorption features would be lost in
our instrumental resolution and sensitivity.  From analysis of models
constructed by adding the N spectrum to a scaled down S spectrum, we
estimate that at most about 15\% of the stars can be in
counterrotation.

A comparison of the measured LOS stellar and [OIII] ionized gas
velocities in the ring shows that the velocities of the stars and the
gas in the polar ring agree closely (see Fig.~\ref{figcompring}).
Note how well the gas velocities bisect the double valued stellar
velocities, indicating that the emission line gas is associated with
the ring, not the S0. We attribute the double valued gas velocities,
$8''<r<15''$, to the different regions sampled by the N and S ring
spectra in the region north of the S0 major axis
(Fig.~\ref{figslitpos}).  The measured H$\alpha$ velocities are
virtually indistinguishable from the [OIII] velocities
(Fig.~\ref{figcompring}); differences are probably due to minor
differences in the slit positions.

Calculating the stellar and emission line rotation curves from the
observed radial velocities requires corrections for inclination, slit
position, asymmetric drift, and LOS integration effects. The
inclination correction is complex because of the warped nature of the
ring (Arnaboldi et al.\ 1997; Gallagher et al.\ 2002). However,
outside of $|\ds0|\sim 6''$ the main part of the polar ring appears
close to edge-on (Gallagher et al.\ 2002), and there the inclination
corrections are negligible.  The correction for the off-center
position of the slit is 7\% at $\ds0=10''$, and less for larger
$\ds0$.  The correction for asymmetric drift depends on the velocity
dispersion and the shape of the velocity ellipsoid, and their
dependence on radius (Binney \& Tremaine 1987, p. 198). In principle these can be derived from the
data presented here, but this is beyond the scope of this Letter. We\break

\hskip-0.3cm\resizebox{0.99\hsize}{!}{\includegraphics{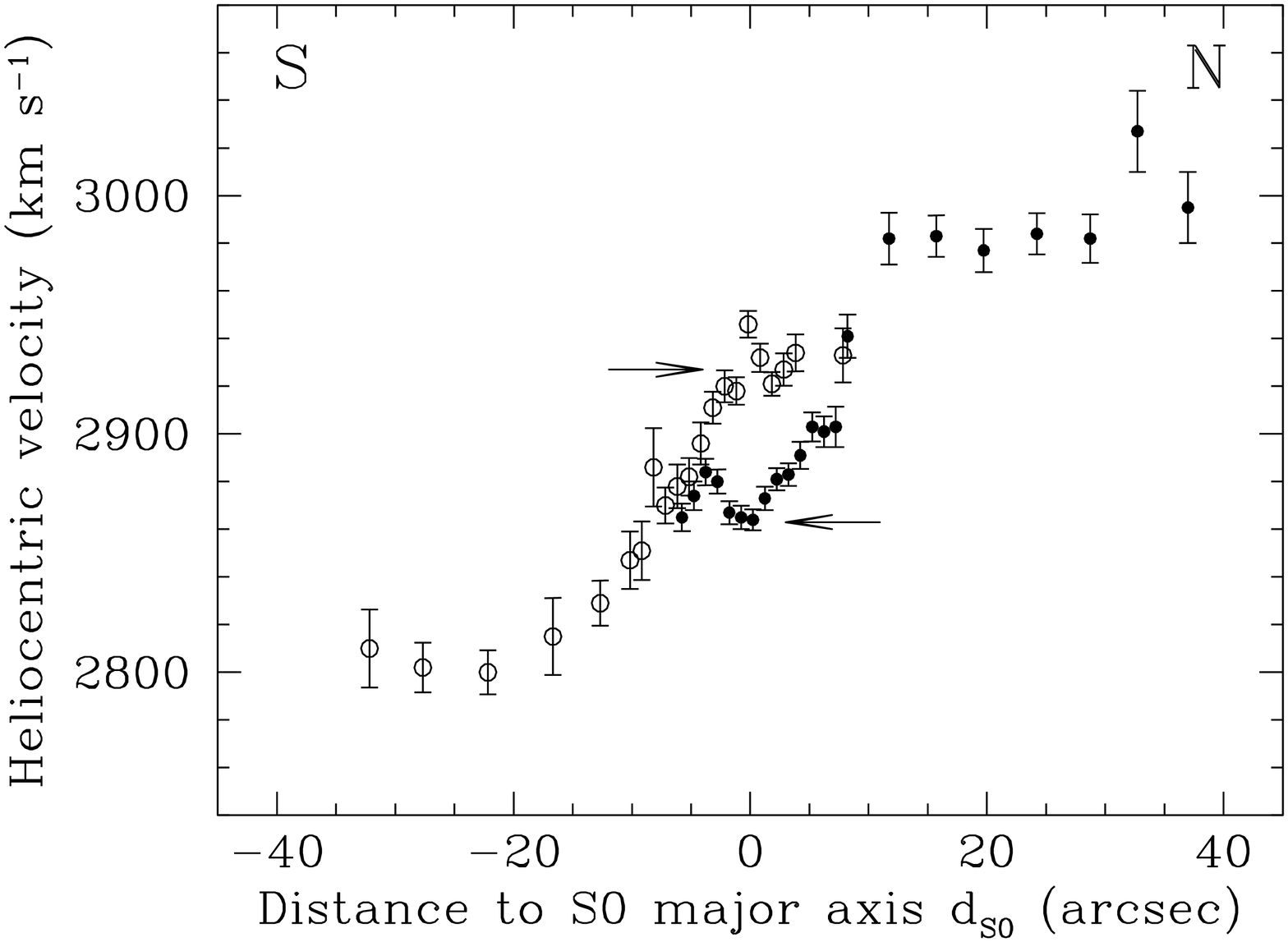}}
{\small {\sc Fig.~\ref{figstellarkin}.}---
Line-of-sight stellar absorption line velocities for the
northern part of the ring (filled circles) and the southern part (open
circles). Each arrow indicates the stellar velocity in the S0 galaxy
(Fig.~\ref{figS0}) at the point where the slit crosses the major axis
of the S0.
}
\bigskip
\addtocounter{figure}{1}

\noindent 
note that the asymmetric drift correction is largest just outside the
galaxy center, and gets smaller with radius.  A final correction is
needed is because the spectra are integrated along the LOS, but these
corrections are small compared to the corrections for asymmetric drift
(Sackett \& Sparke 1990).  Because the corrections are complex or
uncertain, we have not derived the stellar rotation curve for the
polar ring. However, given that the corrections are small, or get
smaller with larger $|\ds0|$, the flatness of the rotation curve is a
secure result.

The agreement between the stellar and emission line kinematics seems
remarkable given that gas and stars are expected to have different
corrections for asymmetric drift and LOS integration. However, after
convolution with the spectral resolution, any narrow spectral features
are washed out, resulting in observed stellar and emission line
profiles that have similar shapes and derived LOS velocities.

LOS stellar velocities along the equatorial plane of the S0
(Fig.~\ref{figS0}), show a smooth rotation curve ($r<20''$ = 3.6
kpc) with LOS velocities approaching 100 km s$^{-1}$.  Emission lines
of [OIII] and H$\beta$ are observed within $\pm 15''$ of the
nucleus. Their LOS velocities, close to the NGC~4650A systemic
velocity, independent of radius, suggest that we are probably
observing the ionized gas component of the polar ring.

\section{Discussion and conclusions}

In this Letter we report the first detection of stellar velocities in
the ring of polar ring galaxy NGC~4650A. To our knowledge, these are
the first stellar velocities obtained for a polar ring. The kinematics
reveal well-defined rotation, both in gas and in stars. The rotation
curve appears to reach a flat part for $|\ds0|>15''$. 

The original impetus for this work was to look for evidence for two
equal mass, counterrotating streams to test the intriguing polar ring
formation model proposed by the Tremaine \& Yu (2000). However, the
fact that we detect well-defined rotation, and the structure of the
CCPs, rule out a polar ring that consists of two equal mass
counterrotating streams.  Nonetheless, given the sensitivity and
resolution of our observations, we cannot rule out that a small
fraction ($\la 15$\%) of the stars is counterrotating.

\hskip-0.3cm\resizebox{0.99\hsize}{!}{\includegraphics{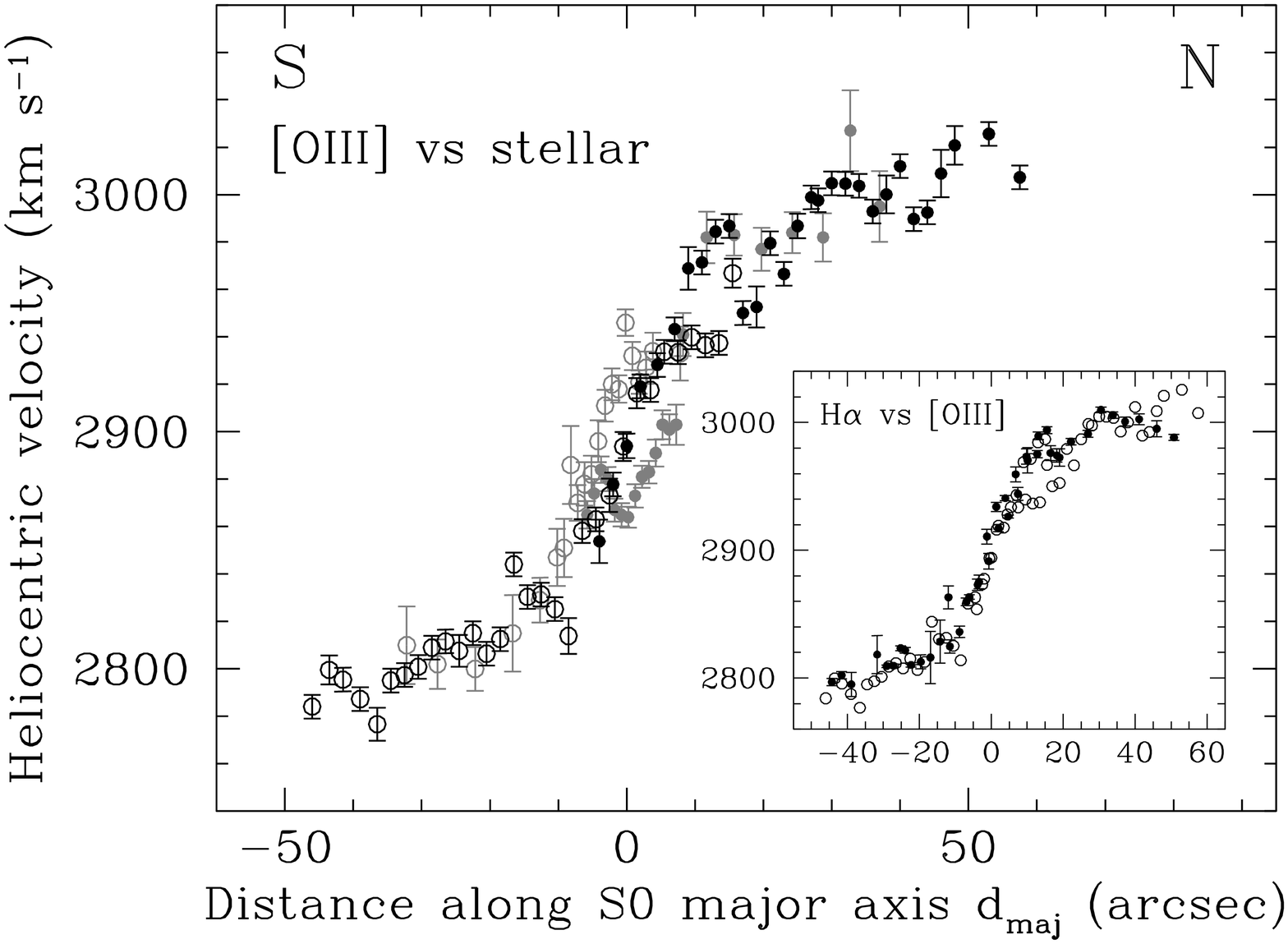}}
{\small {\sc Fig.~\ref{figcompring}.}---
Line-of-sight [OIII] emission line velocities for the N part of the
  ring (filled circles) and the S part (open circles) overlayed on the
  absorption line velocities in gray, with the same coding of the
  symbols. Velocities of the stars and the gas in the polar ring are
  strikingly similar, except in the region $8''<\ds0<15''$, where the
  two slits sample different regions in the ring (see
  Fig.~\ref{figslitpos}). The insert shows a comparison between the
  \Halpha\ (filled circles) and the [OIII] (open circles) velocities.
}
\vskip10pt
\addtocounter{figure}{1}

Thus, it appears that the Tremaine \& Yu (2000) resonance levitation
mechanism is not at work in NGC~4650A. However, as Tremaine \& Yu
(2000) point out, the self-gravity of the ring (likely for NGC 4650A
given its luminous mass is similar to that of the S0, Iodice et
al. 2002) makes it energetically favorable for all stars to be a
single stream, rather than two counterrotating streams.  Another
complication is the uncertain effect of the presence of dust and gas
in the polar ring of NGC~4650A on resonance capture.  Furthermore,
because the resonance levitation model predicts that the polar
material does not extend to the galaxy center, the possibility that
NGC~4650A has a polar disk that does (Iodice et al.\ 2002; see also
below) makes it less likely it has formed through resonance
levitation. Perhaps galaxies with narrow polar rings that have little
or no gas and dust, or disk galaxies with counterrotating streams are
more likely to have formed through resonance levitation.

Absent evidence for a resonance levitation mechanism for the formation
of the polar ring in NGC 4650A, we briefly consider the merger or
acquisition alternatives. The observed flatness of the rotation
velocities in the polar gas and polar stars offers strong kinematic
evidence that this feature is a spatially extended disk rather than a
narrow ring, as suggested earlier from gas kinematics (Arnaboldi et
al. 1997) and from morphology (Gallagher et al. 2002). Insight into
the radial extent of the\break

\hskip-0.3cm\vskip-20pt
\resizebox{0.99\hsize}{!}{\includegraphics{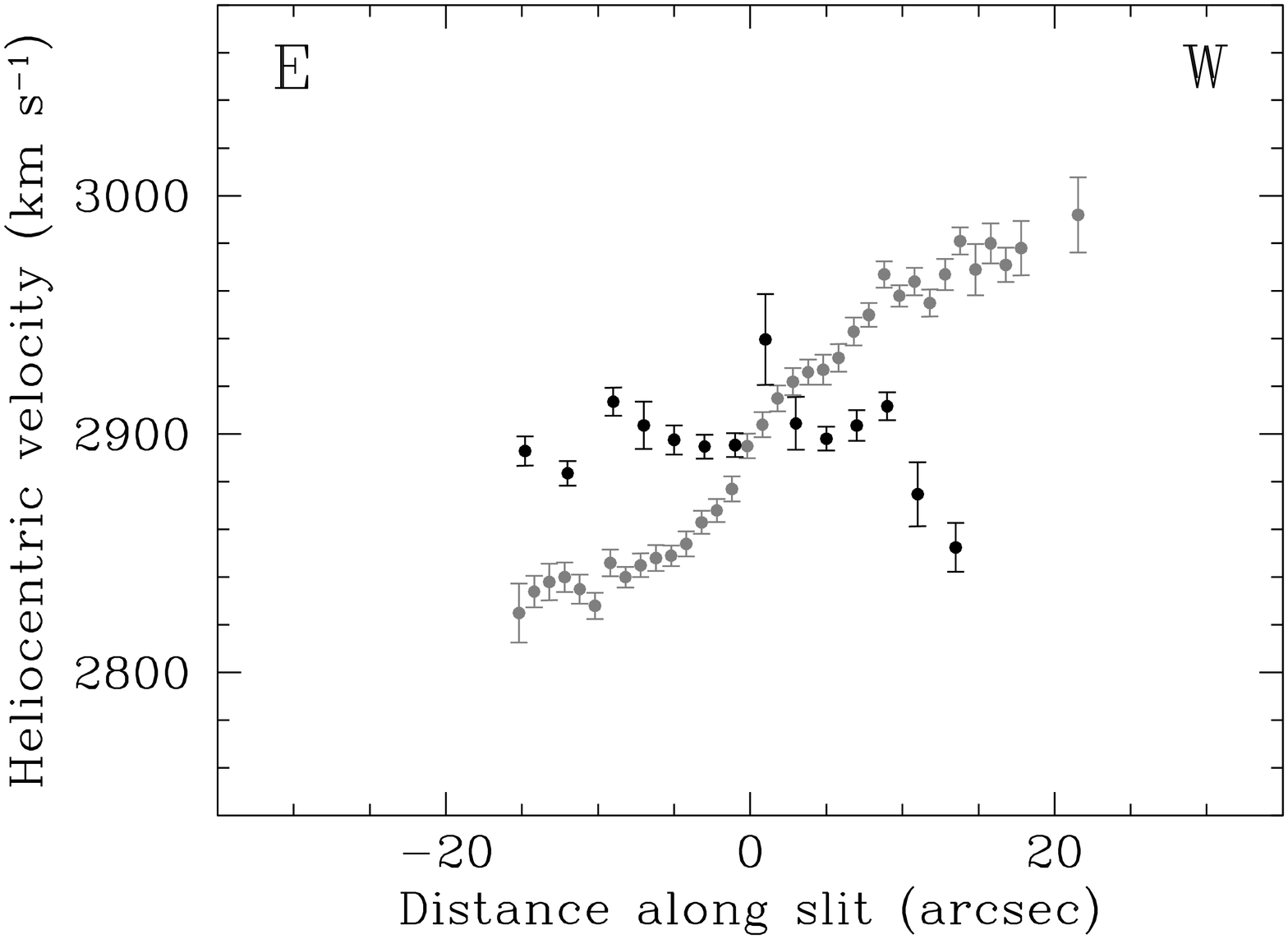}}
{\small {\sc Fig.~\ref{figS0}.}---
Line-of-sight stellar absorption line velocities (gray points) and
  emission line velocities (black points) along the major axis of the
  S0 galaxy. We attribute the emission to the polar ring gas
  component
}
\bigskip
\addtocounter{figure}{1}

\noindent polar ring is offered by the emission line kinematics along the
major axis of the S0.  These velocities, constant at the systemic
velocity, most likely arise from polar ring gas. With the inner ring
warped with an inclination of ~63$^\circ$ (Gallagher et al. 2002), the
emission lines seen at the center indicate that ring gas extends into
the center. The emission likely arises in the chaotic dust lanes which
are oriented at random and extend into the S0 nucleus (Gallagher et
al. 2002).

Iodice et al.\ (2002) argue that the luminous mass in the ring is
similar to or even higher than that of the S0. Its large luminous
mass, combined with its disklike morphology, suggest the polar ring
in NGC 4650A is a disk galaxy that is the victim of a dissipational
polar merger of disks of approximately similar masses (Bekki 1998).
Further support for a merger scenario comes from the location of NGC
4650A in a chain of galaxies in the Centaurus cluster (Sersic
1968). It seems unlikely that a single accretion event could transfer
the required large mass, or that extended accretion over a period of
time could produce sufficient coherence to form a ring.

Although we did not detect counterrotating star streams in the polar
ring of NGC 4650A, which would have been more exciting, we are
continuing the study of this fascinating object. We postpone a
detailed kinematic model until we have more extensive observations.

\acknowledgements

We thank Dr. Maxine Singer, who, as President of the Carnegie
Institution of Washington, insured that the two-Magellan telescope
project proceeded, and the astronomers and technical staff who 
insured that it succeeded.

\end{document}